\documentclass[12pt]{article}
\usepackage{latexsym}
\usepackage{amsmath,,calrsfs}
\usepackage{amsfonts}
\usepackage{amssymb}
\usepackage{amscd}
\usepackage{bbm}
\usepackage{fancybox}
\usepackage{cite}
\usepackage{amsmath,amsfonts,amsbsy}
\usepackage{pstricks,pst-node}
\usepackage[small,bf,hang]{caption2}
\usepackage{graphicx}
\usepackage{epsfig}
\usepackage{psfrag}
\usepackage{comment}
\usepackage{hyperref}

\usepackage{float}

\psset{unit=1.3cm,linewidth=.5pt,radius=.2}  

\usepackage{multirow}                     
\usepackage{float}                          
\usepackage{lscape}                         
\usepackage{bm}

\newcommand{\bb}{\bar\beta}

\newcommand{\beq}{\begin{equation}}
\newcommand{\eeq}{\end{equation}}
\newcommand{\bi}{\begin{itemize}}
\newcommand{\ei}{\end{itemize}}
\newcommand{\bt}{\begin{tabular}}
\newcommand{\et}{\end{tabular}}
\newcommand{\bc}{\begin{center}}
\newcommand{\ec}{\end{center}}

\newcommand{\be}{\begin{equation}}
\newcommand{\ee}{\end{equation}}
\newcommand{\bea}{\begin{eqnarray}}
\newcommand{\eea}{\end{eqnarray}}
\newcommand{\ba}{\begin{array}}
\newcommand{\ea}{\end{array}}

\def\bbox{{\,\lower0.9pt\vbox{\hrule \hbox{\vrule height 0.2 cm
\hskip 0.2 cm \vrule height 0.2 cm}\hrule}\,}}
\newcommand{\dsl}{\pa \kern-0.5em /}

\font\mybb=msbm10 at 12pt
\def\bb#1{\hbox{\mybb#1}}

\def\bE {\bb{E}}

\makeatletter \@addtoreset{equation}{section} \makeatother

\def\slashchar#1{\setbox0=\hbox{$#1$}           
   \dimen0=\wd0                                 
   \setbox1=\hbox{/} \dimen1=\wd1               
   \ifdim\dimen0>\dimen1                        
      \rlap{\hbox to \dimen0{\hfil/\hfil}}      
      #1                                        
   \else                                        
      \rlap{\hbox to \dimen1{\hfil$#1$\hfil}}   
      /                                         
   \fi}

\begin{document}

\begin{titlepage}
\begin{center}

\vskip 1.5cm

{\Large \bf Membranes and gauged supergravity$^*$}

\vskip 1cm

{\bf Paul K.~Townsend} \\

\vskip 25pt

{\em  \hskip -.1truecm
\em  Department of Applied Mathematics and Theoretical Physics,\\ Centre for Mathematical Sciences, University of Cambridge,\\
Wilberforce Road, Cambridge, CB3 0WA, U.K.\vskip 5pt }

{email: {\tt P.K.Townsend@damtp.cam.ac.uk}} \\

\end{center}

\vskip 0.5cm

\begin{center} {\bf ABSTRACT}\\[3ex]
\end{center}

In 1980, Antonio Aurilia, Hermann Nicolai and I constructed an $N=8$ supergravity with a positive exponential potential
for one of the 70 scalar fields by adapting the dimensional reduction of 11D supergravity to allow for a non-zero
4-form field-strength in 4D. This model, now viewed as a particular gauged maximal supergravity, had little influence 
at the time because it has no maximally-symmetric vacuum. However, as shown here, it does have a 
domain-wall solution, which lifts to the M2-brane solution of $D=11$ supergravity. A similar construction 
for other M-branes is also explored.

\vfill

$*$ Contribution to ``Touring the Planck Scale'' - Antonio Aurilia Memorial Volume.

\end{titlepage}

\newpage


\section{Introduction}

Early in 1980, Antonio Aurilia  came to my office at CERN  to tell me about his version, with Christodoulou and Legovini,  of 
the ``bag model'' of hadrons; their bag was the space enclosed by  a membrane coupled to a 3-form  gauge potential such that the 4-form field 
strength was  zero outside the bag but a non-zero constant  inside it \cite{Aurilia:1978qs,Aurilia:1978yc}.  

That was my introduction to the relativistic membrane, but my interest at the time was in the antisymmetric tensor fields that arise in supergravity theories, and 
I was intrigued by the idea that a 3-form potential could be physically relevant even though it has no propagating modes. While listening to Antonio, it
occurred to me that a constant 4-form  field strength would  be a cosmological constant in a gravitational context. We soon worked out  the details, in which 
the cosmological constant emerges as an integration constant in the solution of the field equation for the 3-form potential\footnote{As was independently 
discovered around the same time by Duff and van Nieuwenhuizen \cite{Duff:1980qv}, and (in the context of a superspace formulation of $N=1$ supergravity) by Ogievetsky and Sokatchev 
\cite{Ogievetsky:1980qp}.}, and we then had the idea of applying the result to 11D supergravity \cite{Cremmer:1978km}; 
the 3-form gauge potential of that theory implies, in the context  of dimensional reduction,  a 3-form gauge potential for 4D $N=8$ supergravity. 
Cremmer and  Julia had recently found the full $N=8$ supergravity action in this way \cite{Cremmer:1978ds},  but they had  set to zero the 4D 4-form field strength.
Working with Herman Nicolai,  we found a more general $N=8$ supergravity action with an exponential  scalar potential for one of the  70 scalar 
fields \cite{Aurilia:1980xj}.  The  cosmological constant had effectively been traded  for the expectation value of a scalar field; this was 
interesting but the absence of any maximally  symmetric 4D vacuum was disappointing\footnote{We addressed this issue briefly
in a conference report \cite{Aurilia:1980xy}, where it was observed that a positive potential is what one might expect from a spontaneous partial breaking of 
the local supersymmetry.}.  

The idea that a 3-form potential could replace a cosmological constant soon attracted attention.  An example that is noteworthy here, because it has elements
in common with the  Aurilia-Chistodoulou-Legovini bag model,  is the 1987 work  of Brown and  Teitelboim \cite{Brown:1987dd} in which it is  shown that the 
cosmological constant, interpreted as the magnitude $F$ of a 4-form field strength, could be dynamically reduced by nucleation of membrane  bubbles if  
$F$ is  lower inside the bubble than outside.  In contrast, our new $N=8$ supergravity had very little impact,  
presumably because it was not related to other ideas of the time, and thus did not appear to be part of some bigger picture. 

The bigger picture began to emerge a few months later, when Freund and Rubin showed that 11D supergravity has an $AdS_4\times S^7$ solution with the
 4-form field strength proportional to the volume form of $adS_4$ \cite{Freund:1980xh}.  It was soon conjectured, and later proved, that the associated 4D theory is a gauged 
 maximal supergravity  with an  $SO(8)$ gauge group, which was constructed by de Wit and Nicolai  in 1982 \cite{deWit:1981sst}. These developments were reviewed 
in 1986  by Duff, Nilsson and Pope \cite{Duff:1986hr}; by then it was understood that  the de Wit-Nicolai  theory is but one of many gauged 
maximal supergravity theories. A classification was achieved in relatively recent times; this was reviewed in 2008 by Samtleben \cite{Samtleben:2008pe}, 
who points out that the  modified $N=8$ supergravity of  \cite{Aurilia:1980xj} is included.  In other words, the $N=8$ supergravity found in 1980 by Antonio, 
Hermann and myself was a gauged maximal supergravity theory {\it avant la lettre}. 

The aim of this article is to show, in a different way, how this first gauged maximal supergravity fits into the bigger picture of M-theory.
Although it has no Minkowski vacuum, it does have  a membrane/domain-wall `vacuum' solution that preserves half the supersymmetry.  Considering that the 
starting point of my collaboration with Antonio was his idea that a relativistic membrane is a source for a 3-form gauge potential\footnote{To my knowledge, this idea originated 
in Antonio's 1977 paper with Legovini \cite{Aurilia:1977jz};  it is, of course, a straightforward generalization of the Kalb-Ramond coupling of strings 
to a 2-form potential \cite{Kalb:1974yc}.}, it now seems surprising that we  did not immediately look for a domain-wall solution of our new
$N=8$ supergravity theory.  Of course, there was then no understanding of how  a membrane coupling to the 3-form of 11D supergravity
could be compatible with local supersymmetry; the  11D supermembrane lay seven years in the future \cite{Bergshoeff:1987cm}.  However, if we had 
looked for, and found, the half-supersymmetric  4D domain wall solution, it would surely have been obvious that this must lift  to a membrane solution 
of 11D supergravity.   In fact, as will be shown here, it lifts to the  the  membrane solution  found by Duff and Stelle in 1990 \cite{Duff:1990xz}. 

I will begin with an exposition of the construction of \cite{Aurilia:1980xj}, which I will refer to as the ``ANT construction'', after its authors. 
In contrast to the detailed exposition there, no attempt will be made here to include all supergravity fields, including fermions. Instead, I will start with a simplified
model of gravity in a spacetime of general dimension $D=d+n$, coupled to a $d$-form field strength $F_d$. This suffices for an explanation of the basic idea,
which yields (in this simplified but also generalized context) a $d$-dimensional dilaton-gravity model; for $(d,n)=(4,7)$ it is 
a consistent truncation of the modified $N=8$ supergravity theory found in \cite{Aurilia:1980xj}.  For certain other values of $(d,n)$ a similar construction may 
apply for other M-theory branes, as will be discussed.

\section{The ANT construction}

Our starting point will be the following Lagrangian density 
\begin{equation}\label{start}
{\cal L}_D = \sqrt{-g^{(D)}}\left\{R^{(D)} - \frac12 |F_d|^2\right\}\, , 
\end{equation}
where $|F_d|^2$ is defined such that, in local coordinates for which the $D$-metric is $g_{MN}$ and $F_d$ has components $F_{M_1\cdots M_d}$, 
\begin{equation}
|F_d|^2 = \frac{1}{d!} g^{M_1N_1} \cdots g^{M_dN_d} F_{M_1\dots M_d}F_{N_1\dots N_d} \, .
\end{equation}
We will be interested in an $n$-dimensional reduction  to a $d$-dimensional spacetime ${\cal M}_d$  (so $D=d+n$) with a reduction/truncation ansatz for which 
\begin{eqnarray}\label{Dansatz}
ds^2_D &=&  e^{-2\alpha\phi} ds^2({\cal M}_d)  + e^{2\left[(d-2)/n\right]\alpha\phi} ds^2(\bE^n) \nonumber\\
F_d &=& {\rm vol}\left({\cal M}_d\right) f\, , 
\end{eqnarray}
where ${\rm vol}({\cal M}_d)$ is the volume $d$-form on ${\cal M}_d$, and $\alpha$ is an arbitrary constant.  Although both $\phi$ and $f$ are scalar fields on  ${\cal M}_d$ that are constant on the Euclidean $n$-space $\bE^n$,  the scalar $f$ is constrained  to be the Hodge dual of a $d$-form field strength on ${\cal M}_d$. 

The metric ansatz has been chosen such that $ds^2({\cal M}_d)$ is the  Einstein conformal frame metric. A choice of the constant $\alpha$ is equivalent to a normalization for $\phi$, and if  we choose
\begin{equation}\label{alpha}
\alpha = \sqrt{\frac{n}{2(d-2)(D-2)}}\, , 
\end{equation}
then the Lagrangian density for the $d$-dimensional theory is 
\begin{equation}\label{Lagd}
{\cal L}_d = \sqrt{-g}\left\{R  - \frac12 (\partial\phi)^2 + \frac12 e^{2(d-1)\alpha\phi} f^2\right\}\, . 
\end{equation}
For $(d,n)=(4,7)$ this is  a consistent truncation of the full $N=8$ supergravity found by dimensional reduction of 11D supergravity.  If the scalar $f$ were unconstrained, its equation of motion would 
be $f=0$; we could then back substitute to trivially eliminate $f$ from the action, which would then be, for $(d,n)=(4,7)$, a  consistent truncation of the standard Cremmer-Julia $N=8$ supergravity. 

However, $f$ is not unconstrained;  the unconstrained variable is the $(d-1)$-form potential for (the Hodge dual of) $f$ and {\it its} field equation 
has the general solution
\begin{equation}\label{f}
f = \mu \,  e^{-2(d-1)\alpha\phi}\, , 
\end{equation}
for arbitrary mass parameter $\mu$. For $\mu\ne0$ it is not legitimate to substitute for $f$ into the Lagrangian density of (\ref{Lagd}) because
the variation of the action is not zero for this solution of the field equations. This can be remedied by first adding to the Lagrangian density a  $\mu f$ term,  which is 
a total derivative because of the constraint on $f$. It is now legitimate to substitute for $f$, and this  yields the new `dual' Lagrangian density
\begin{equation}\label{tLagd}
\tilde {\cal L}_d = \sqrt{-g}\left\{R  - \frac12 (\partial\phi)^2 - 2\Lambda e^{-a\phi} \right\}\, , 
\end{equation}
where
\begin{equation}
a= 2(d-1)\alpha\, , \qquad \Lambda = (\mu/2)^2\, . 
\end{equation}
Notice that the sign of the potential term is {\it opposite} to what one would find by the {\it illegitimate} substitution for $f$ in (\ref{Lagd}). This sign change
may be checked by verifying that the field equations of (\ref{Lagd}) are equivalent, for $f$ given by (\ref{f}), to the field equations of (\ref{tLagd}). 

Our conventions are now those of \cite{Lu:1996rhb} except that $\Lambda$ here is $-\Lambda$ there;  defined here, $\Lambda$ becomes 
the usual cosmological constant (negative for anti-de Sitter) when $a=0$.  Let us record here that, as a consequence of (\ref{alpha}), 
\begin{equation}\label{asquared}
a^2 = \frac{2n(d-1)^2}{(d-2)(D-2)}\, .  
\end{equation}

\section{Domain walls and M-branes}

The domain-wall solutions  of the field equations that follow from the Lagrangian density  of (\ref{tLagd}) take the form \cite{Lu:1995cs,Lu:1996rhb}
\begin{eqnarray}\label{dwd}
ds^2_d &=& H^{\frac{4}{(d-2)\Delta}} ds^2({\rm Mink}_{d-1}) + H^{\frac{4(d-1)}{(d-2)\Delta}} dy^2 \nonumber\\
e^\phi &=& H^{\frac{2a}{\Delta}}\, , 
\end{eqnarray}
where $H$ is linear in $y$ with $dH= \mp \sqrt{\Lambda\Delta}\, dy$, and 
\begin{equation}
\Delta \equiv  a^2 - \frac{2(d-1)}{(d-2)} \, . 
\end{equation}
From (\ref{asquared}) we see that, for us, 
\begin{equation}\label{Delt}
\Delta =  \frac{2(n-1)(d-1)}{d+n -2} \, . 
\end{equation}

Our aim now is to lift these domain wall solutions to solutions of the  $D$-dimensional theory from which we started, 
using the ansatz (\ref{Dansatz}). The spacetime ${\cal M}_d$ appearing in this ansatz is now the above domain-wall 
spacetime, and 
\begin{equation}
 {\rm vol}\left({\cal M}_d\right) = H^{\frac{4(d-1)}{(d-2)\Delta}} {\rm vol}({\rm Mink}_{d-1}) \wedge dy \, . 
 \end{equation}
 Let us also record here that for the domain-wall solution, the function $H(y)$ is such that 
 \begin{equation}
dH^{-1} = \pm \sqrt{\Lambda\Delta}\, H^{-2}  dy \, . 
\end{equation}

On substituting the domain-wall metric and dilaton configurations of (\ref{dwd})  into the ansatz (\ref{Dansatz}), one finds that 
\begin{eqnarray}\label{M2}
ds^2_D &=& H^{-\frac{2}{d-1}} ds^2({\rm Mink}_{d-1}) + H^{\frac{2}{n-1}} ds^2 (\bE^{n+1}) \nonumber \\
F_d&=& \pm \frac{2}{\sqrt{\Delta}} \,  {\rm vol}({\rm Mink}_{d-1}) \wedge dH^{-1}\, . 
\end{eqnarray}
We have been assuming that $H$ is a function only of $y$, but this is now a special case of a more general $(d-1)$-brane solution of the 
the $D$-dimensional theory from which we started; in general $H$ is a harmonic function on the $(n+1)$-dimensional transverse Euclidean space. 
A simple choice, with $SO(n+1)$ symmetry, is 
\begin{equation}
H= 1+ (r_0/r)^{n-1}\, , 
\end{equation}
for constant $r_0$.  Near the singularity of  this function at $r=0$, the $D$-metric takes the asymptotic form 
\begin{equation}
H \sim e^{-(n-1)\rho/r_0} ds^2({\rm Mink}_{d-1})  + d\rho^2 + r_0^2 d\Omega_n\qquad \left[\rho=r_0 \ln (r/r_0) \right], 
\end{equation}
where $d\Omega_n$ is the $SO(n+1)$-invariant metric on $S^n$. This is an AdS$_d \times S^n$ metric, which implies that $r=0$ is a
Killing horizon of the full $D$-metric, near which the solution asymptotes to AdS$_d \times S^n$. 

The configuration (\ref{M2}) is a solution of a $D$-dimensional supergravity in those cases of  table \ref{tbl:branesdpw}, which is adapted from a similar 
table in  \cite{Claus:1998mw}. The first column gives the type of solution. The 
second column gives the spacetime dimension $D$ and the next two columns gives the values of $d$ and $n$. The last column
gives the value of $\Delta$ according to the formula (\ref{Delt}). 
\begin{table}[h]\begin{center}\begin{tabular}{||l|c|c|c||c||}\hline
    {\rm Solution \ type}                          & D & d & n  & $\Delta$   \\ \hline
M2                            & 11& 4 & 7 & 4  \\
M5                            & 11& 7 & 4 & 4    \\ \hline
D3                            & 10& 5 & 5 & 4    \\
\hline\hline
Self-dual black string              & 6 & 3 & 3 & 2    \\\hline
Magnetic black string               & 5 & 3 & 2 & 4/3   \\
Tangerlini black hole         & 5 & 2 & 3 & 4/3  \\\hline
Reissner-Nordstr\"om black hole & 4 & 2 & 2 & 1   \\
\hline
\end{tabular}\end{center}\caption{Supergravity solutions with an `AdS$\times$S' near-horizon  geometry.}
\label{tbl:branesdpw}\end{table} 
Notice that entries with $d\ne n$ occur in `dual' pairs for which the values of $d$ and $n$ are interchanged; these have the
same value of $\Delta$, as is manifest from the formula (\ref{Delt}).   

For $(d,n)=(4,7)$, in which case $\Delta=4$, we  recover  the half-supersymmetric 
Duff-Stelle membrane solution of 11D supergravity \cite{Duff:1990xz} as a lift to $D=11$ of a domain-wall solution of  the modified $N=8$ supergravity  
theory of \cite{Aurilia:1980xj}.

\subsection{Other cases}

What about the other cases of the table? For $(d,n)=(7,4)$, we recover the half-supersymmetric  fivebrane solution of 11D supergravity, but in terms of a 
7-form field strength for the 6-form `dual' potential that is defined only on solutions of the 11D supergravity equations. This  is probably only a technical difficulty since the ANT construction could
be recast as a purely on-shell construction, but this will not be attempted here.  The remaining $\Delta=4$ case involves the self-dual 5-form field strength  of  IIB 10D supergravity, 
and we now face the complication that  a non-zero 5-form on the 5-dimensional spacetime  of the maximal 5D supergravity obtained by dimensional reduction  must be accompanied by a non-zero 5-form field strength on the Euclidean 5-space on which we reduce; again, some modification of the ANT construction may allow for this  but this will not be investigated here either. 

What about the $D<10$ cases with $\Delta<4$? The are various intersecting M-brane solutions of 11D supergravity for which the asymptotic geometry near the intersection
is of `AdS $\times S \times \bE$' form. The four geometries of this type obtainable in this way, with a particular realization in terms of intersecting M2-branes and M5-branes
are given in table \ref{tbl:intersections}, which is adapted from a similar table in \cite{Boonstra:1998yu}. The symbol $\perp$ indicates either (i) an orthogonal intersection 
in which p-branes self-intersect on $(p-2)$-planes, which may then also intersect on $(p-4)$-planes, or (ii) the orthogonal intersection of an M2 with an M5 on a line; these
intersections are among those that preserve some fraction of supersymmetry and the fraction preserved is given in the last column of the table. 
\begin{table}[h]
\begin{center}
\begin{tabular}{|c|c||c||}
\hline
$M2\perp M5$ & $adS_3\times S^3\times \bE^5$ & 1/4 \\ \hline
$M2\perp M2\perp M2$ & $adS_2\times S^3\times \bE^6$ & 1/8\\ \hline
$M5\perp M5\perp M5$ & $adS_3\times S^2\times \bE^6$ & 1/8\\ \hline
$M2\perp M2\perp M5\perp M5$ & $adS_2\times S^2 \times \bE^7$& 1/8\\
\hline
\end{tabular}\end{center}
\caption{{\it Intersecting M-branes with `AdS $\times S \times \bE$' near-horizon geometries.}} 
\label{tbl:intersections}
\end{table}

Consider the 6D self-dual black string solution of the minimal $(1,0)$ 6D supergravity, which is a consistent truncation of the maximal 6D supergravity obtained by dimensional reduction 
of 11D supergravity.  An application of the ANT construction would, if successful, lead to a $d=3$ $N=4$ supergravity with a domain-wall vacuum that lifts to the 6D self-dual string.
However, the 3-form field strength that supports this solution is self-dual, and this presents the same difficulty that we found for the D3 case. Next is the magnetic black string: a 
half-supersymmetric solution of minimal 5D supergravity, which is a consistent truncation of the maximal 5D supergravity obtained by dimensional reduction of 11D supergravity. 
Here we have the same issue that arose for the M5 case, the 3-form field strength that we need for an application of the ANT construction is defined only on-shell. These 
difficulties (which may be purely technical)  do not arise for the remaining, extreme black hole, cases but for these the ANT construction would require us to dimensionally reduce to 
2D, which is the dimension for which  there is no Einstein-frame metric. 

So, the ANT construction of \cite{Aurilia:1980xj} leading to a modified $N=8$ supergravity with a domain-wall `vacuum' that lifts to the M2-brane of 11D supergravity
does not immediately generalise to the other M-theory possibilities, although a modified construction may work in some of these other cases.

\section{Comments}

My interaction with Antonio Aurilia was brief but intense.  Apart from our joint work with Hermann Nicolai, we also co-authored a paper
with Antonio's  long-term (and long distance) collaborator Yasushi Takahashi (of Ward-Takahashi fame)  on another application of
a non-propagating 3-form gauge potential \cite{Aurilia:1980jz}. I eventually met my co-author Takahashi at the Edmonton Summer 
Institute of 1987, although he was speaking at a parallel event on superconductivity\footnote{My talk at the other event was on branes, as was that of Kellogg Stelle with whom 
I co-wrote a conference proceedings with the rhetorical  title {\it Are 2 branes better than 1} \cite{Stelle:1987bi}.};  I introduced myself to him after his talk, mentioning our joint paper but he had no recollection of it! 

I think Antonio and I  met only once again after our 1980 collaboration at CERN,  but his influence 
on me was very significant. I paid attention to his ideas about membranes at the same time that I was working with him 
on 11D supergravity, and I returned to wondering how to connect these topics  after the Green-Schwarz  superstring revolution of 1984. Eventually, 
in Trieste in January 1987, Eric Bergshoeff, Ergin Sezgin and I were able to solve the problem of how to couple 11D supergravity to a
membrane \cite{Bergshoeff:1987cm}. This set the course for much of my subsequent research, although branes only became mainstream in 1996. 
Before then, Antonio was a fellow heretic; see, for example,  his 1993 paper with Spallucci \cite{Aurilia:1993xi}. 
He would have been thrilled to know that our joint work on 11D supergravity would also turn out to be  related to  the topic so close to his heart -- 
the relativistic membrane.

\subsection*{Acknowledgements}

Partial support from the UK Science and Technology Facilities Council (consolidated grant ST/P000681/1) is gratefully acknowledged.


\providecommand{\href}[2]{#2}\begingroup\raggedright\endgroup

\end{document}